\documentclass[12pt,reqno]{article}
\usepackage{amsmath,epsfig}
\usepackage{amssymb,amsfonts}
\usepackage{latexsym}
\usepackage{graphicx}
\usepackage[a4paper]{geometry}
\usepackage{euscript}%

\usepackage{slashed}
\usepackage{dsfont}

\usepackage[margin=10pt,font=small]{caption}

\relax

\def\be{\begin{equation}}
\def\ee{\end{equation}}
\def\bea{\begin{eqnarray}}
\def\eea{\end{eqnarray}}

\newcommand\fverb{\setbox\pippobox=\hbox\bgroup\verb}
\newcommand\fverbdo{\egroup\medskip\noindent%
                        \fbox{\unhbox\pippobox}\ }
\newcommand\fverbit{\egroup\item[\fbox{\unhbox\pippobox}]}

\newcommand{\bear}{\begin{eqnarray}}

\newcommand{\eear}{\end{eqnarray}}

\newcommand{\bse}{\begin{subequations}}
\newcommand{\ese}{\end{subequations}}
\newcommand{\ea}{\end{array}}
\newbox\pippobox

\def\lab{\label}
\def\6{\partial}
\def\f{\phi}
\def\a{\alpha}

\def\le{\left}
\def\ri{\right}
\def\cO{{\cal O}}

\def\cL{{\cal L}}
\def\C0{{\bf C_0}}
\def\Y0{{\bf Y_0}}
\def\G0{{\bf G_0}}

\def\r{\rho}

\def\sq
\def\a{\alpha}
\def\b{\beta}
\def\l{\lambda}
\def\g{\gamma}

\def\eps{\epsilon}

\def\o{\omega}

\def\bz{\begin{itemize}}
\def\ez{\end{itemize}}
\def\bn{\begin{enumerate}}
\def\en{\end{enumerate}}

\def\ben{\begin{enumerate}}
\def\een{\end{enumerate}}

\def\q{\theta}

\allowdisplaybreaks[2]
\numberwithin{equation}{section}

\newcommand{\ov}[1]{\overline{#1}}

\def\a{\alpha}
\def\o{\omega}
\def\6{\partial}
\def\lab{\label}

\def\le{\left}
\def\ri{\right}

\def\be{\begin{equation}}
\def\ee{\end{equation}}
\def\bea{\begin{eqnarray}}
\def\eea{\end{eqnarray}}
\def\bz{\begin{itemize}}
\def\ez{\end{itemize}}

\def\cL{{\cal L}}
\def\cO{{\cal O}}

\begin{document}
\begin{flushright} \small
 IPM/P-2011/044 \\ CERN-PH-TH-2011-285\\
 NSF-KITP-11-219
\end{flushright}
\bigskip
\begin{center}
 {\large\bfseries Rotating strings and energy loss in non-conformal holography}\\[5mm]
Mohammad Ali-Akbari$^*$,  Umut G\"ursoy$^{\dag}$,  \\[3mm]
 {\small\slshape
 *  School of Physics, Institute for Research in Fundamental Sciences (IPM), P.O.Box 19395-5531, Tehran, Iran
\\
 \medskip
  \dag Theory Group, Physics Department, CERN, CH-1211 Geneva 23, Switzerland\\
\medskip
 {\upshape\ttfamily aliakbari@theory.ipm.ac.ir; umut.gursoy@cern.ch}\\[3mm]}
\end{center}
\vspace{5mm} \hrule\bigskip \centerline{\bfseries Abstract}
We study the energy lost by an accelerating quark probe in the quark-gluon plasma produced in the heavy ion collisions in an approximate setting where the acceleration of the probe is due to uniform circular motion. The energy loss rate of the rotating probe is calculated at strong coupling in the confining $SU(N)$ gauge theory based on N $D4$ branes on a circle, using the rotating string solutions in the dual gravitational background. The system is known to exhibit a confinement-deconfinement transition at a finite temperature $T_c$. We investigate energy loss both in the low and the high T phases. The high T phase is similar to the previously studied case of the ${\cal N}=4$ plasma, yet we find differences due to non-conformality of the underlying theory. The low T phase, on the other hand exhibits novel interesting behavior: We find a dual gravitational mechanism that yields a lower bound on the emitted energy of the rotating quark, proportional to the mass gap in the glueball spectrum. The low T energy loss is argued to be completely due to glueball brehmstrahlung, hence the energy loss rate calculated here determines the Lienard potential for syncrotron radiation in this confining gauge theory at strong coupling.
\medskip

\bigskip
\hrule\bigskip
\newpage
\section{Introduction}\lab{sec:intro}

Energetic partons traversing a hot plasma provide very important observables in the heavy ion
collisions (for a recent review with emphasis on holographic methods
see  \cite{review}), that exhibit distinctive properties such as jet-quenching which can clearly be observed at RHIC \cite{RHIC} and more recently at LHC \cite{LHC}. Depending on the energy of the projectile, the energy loss can be due to various different mechanisms such as the destructive interference between vacuum radiation and QCD brehmsstrachlung or elastic scattering of the projectile with the surrounding medium. Theoretical study of these phenomena is notoriously difficult as the the  perturbative QCD calculations fall short, and one needs strong-coupling input at various stages of the computations. Non-perturbative methods, such as lattice QCD are
also inadequate to describe such time-dependent phenomena.

This provides the main motivation behind the energy loss studies using gauge-gravity duality
\cite{Gubser1}\cite{Seattle}\cite{LRW}\cite{CST}. In approach of \cite{Gubser1,Seattle} the dominant
mechanism is assumed to be the elastic interactions of the parton with plasma. These interactions are further assumed to be strongly
coupled at all relevant energy scales. The simplest setting involves a quark traveling with constant velocity in linear motion throughout the plasma, and the question is the energy required to keep it in uniform motion using the AdS/CFT correspondence\cite{AdSCFT}.   On the gravity side, the quark is represented by the end-point of a string at the boundary. The string that trails
the quark extends toward the interior of the dual black-hole geometry, reaching the horizon. The energy required to keep the quark in uniform motion is then given by the world-sheet momentum that falls across the horizon.
In the case of conformal plasmas such as the ${\cal N}=4$ super Yang-Mills theory, this energy turns out to be proportional to the momentum of the parton, thus energy loss happens via drag. In more realistic examples that imitate the energy dependence of the strong interactions \cite{GK,GKN}, one finds that the drag coefficient is not constant
but it also depends on the momentum of the quark \cite{GKMiN,GKMN4}.

In the real setting of the heavy-ion experiments however there is no external force acting on the parton keeping it in
constant velocity. Rather, the parton decelerates. On the other hand,
it turns out a challenging problem to study linearly  decelerating
projectiles in the gauge-gravity duality. A technically more tractable
case is when the quark is in uniform rotation with constant angular
velocity, where one can construct the relevant rotating string
solution relatively easily. Motivated by this, the authors of
\cite{Fadafan} studied this problem in case of the conformal ${\cal
  N}=4$ plasma at finite temperature and strong coupling. Generally,
 energy loss is a combination of two effects: the drag force due to
interaction with the medium and the syncrotron radiation of the
rotating probe. The drag effect becomes dominant in the regime $\o\to
0$, $L\to\infty$, with constant velocity $v = \o L $, where the
calculation \cite{Fadafan} reproduces the earlier result of the linear
drag force \cite{Gubser1}. In the opposite, high frequency and small
radius limit $\o\to\infty$, $L\to 0$,  $v= \o L $ constant, the latter effect becomes dominant and the calculation
reproduces the Lienard potential for syncrotron radiation
\cite{Mikhailov}.

In this paper we extend this study to non-conformal plasmas, where the vacuum of the theory at zero temperature confines the color change. As a first step in this direction, first we would like to understand the qualitative differences that arise due to non-conformality and confinement. In particular, these theories possess a fundamental energy scale that we refer to as $\Lambda_{QCD}$ in addition to temperature\footnote{Of course, presence of a fundamental energy scale does not necessarily imply confinement. We refer to \cite{GKN} for a classification of confining theories in a general 5D holographic setting.}. The rate of energy loss should therefore be qualitatively different at energy scales smaller or bigger than $\Lambda_{QCD}$. More precisely, the theory we consider exhibits a confinement-deconfinement transition at some finite temperature $T_c \propto \Lambda_{QCD}$, below which the fundamental excitations are color-blind objects such as the glueballs\footnote{In this paper we only consider pure Yang-Mills coupled to operators in the adjoint representation of the gauge group, therefore the only hadrons are glueballs.}. Above the transition, the color charge deconfines and medium becomes a gluon plasma. Clearly,
energy loss of a rotating probe in the low T phase should only be due to syncrotron radiation, whereas in the high T phase both radiation and drag may play a role.

We study the problem in the celebrated model of Witten \cite{Witten1} based on N D4 branes wrapped on a circle. The field theory is a non-supersymmetric $SU(N)$ Yang-Mills theory that confines at low temperatures \cite{Witten1}. Although the theory contains infinite number of undesired scalar operators coupled to the glue sector, thus very different than pure Yang-Mills theory, it is argued to be in the same universality class \cite{Witten1} in the sense that it exhibits linear confinement of quarks in the vacuum. What makes this model attractive for our purposes here is two-fold. Firstly, it is a top-down approach where one can control the stringy corrections parametrically; secondly, the gravitational backgrounds dual to both the vacuum and the high T phase is known analytically; a fact that simplifies the calculations substantially.

In the next section we first review the calculation of energy-loss of rotating quarks in the dual gravitational setting. We keep the discussion as general as possible, and present formulae for the energy-loss rate that can be applied to a large class of backgrounds. Then we review the features of the $D4/S^1$ model that will be used in the following sections. The third section is devoted to the study of rotating probes in the low T phase. In particular we show that there is general mechanism on the gravitational set-up that yields a lower bound on the possible energy loss by a rotating quark, that is given by the mass gap of the gluons. We further argue that the energy loss in the low T phase is completely due to syncrotron radiation. Therefore, our calculation provides the Lienard potential for the syncrotron radiation at strong coupling for the confining gauge theory living on $D4/S^1$ branes. The fourth section studies rotating probes in the high T phase. We end the paper with a discussion section where we summarize our results and point towards future directions.

\section{Review of background material}

\subsection{Generalities of rotating strings and energy loss}
\lab{rev1}

 We begin by reviewing general features of rotating string solutions in generic string backgrounds and its relation to energy loss of rotating probes in the dual field theory.
The ansatz for a rotating string is given by\footnote{All other coordinates in possible internal dimensions are fixed to be constant.},
\be \lab{ansatz}
 X^M\equiv(t=\tau, u=\sigma, \phi=\o t+\theta(u), \rho=\rho(u), x_3=0).
\ee %
Here $\tau$ and $\sigma$ are the world-sheet coordinates and $u$ is the radial coordinate in the space-time geometry that is typically a domain-wall with translation invariance in the 3+1 Minkowski directions. The boundary of the space-time is at $u=\infty$. We choose a radial parametrization of two the Minkowski directions $t,x_i$ as $x_1= \r \cos(\q)$ and $x_1= \r \sin(\q)$.

The first two entries in (\ref{ansatz}) corresponds to the static gauge choice for the world-sheet parametrizations.
The third entry describes a string rotating with angular velocity $\o$ and and off-set angle $\q$ that depends on the radial variable $u$. $\r$ is the radius of rotation at a given plane $u=const.$

The end of the string corresponds to the quark rotating the in the plasma. This end point is rotating in the $(x^1,x^2)$ plane with an angular frequency $\o$. Radius of its circular motion is a parameter of the problem, that we denote by $L$. Thus, the boundary conditions for the string solution at $u=\infty$ are\footnote{The latter is just a convenient choice with no loss of generality.}:
\be\lab{bc}
 \rho(\infty) = L, \qquad  \q(\infty) = 0.
 \ee
Shape of the string is determined by the extremum of the Nambu-Goto action\footnote{In principle one has to also consider coupling of the string to dilaton in the form $\int d\tau d\sigma \phi R^{(2)}$ where $R^{(2)}$ is the world-sheet curvature. This term is $\cO(\ell_s^2)$ suppressed w.r.t the area coupling however, hence it is safe only to consider the NG action in the SG limit.},
\be\lab{NG}
 S=-\frac{1}{2\pi\alpha'}\int d\tau d\sigma{\cal{L}}
 =-\frac{1}{2\pi\alpha'}\int d\tau d\sigma\sqrt{-\det g_{\mu\nu}}
\ee %
where $g_{\mu\nu}=G_{MN}\partial_\mu X^M \partial_\nu X^N$ is the induced metric on the world-sheet. It is given by
\be\lab{wsmet}
g_{\a\b} = \le(\begin{array}{cc} G_{tt} + G_{\f\f} \o^2 & G_{\f\f} \o \q' \cr G_{\f\f} \o \q'  & G_{uu} + G_{\r\r} \r^{'2} + G_{\f\f} \q^{'2}  \end{array} \ri)
\ee
The solution is parametrized by two functions $\rho(u)$ and $\q(u)$. A general feature of the string solutions in this paper is independence of the NG Lagrangian (\ref{NG}) of the angular variable $\q$. This is because the angle $\phi$ only appears under a derivative in (\ref{NG}) because the space-time geometries we consider in this paper (see eqs. (\ref{lowT}) and (\ref{highT})), do not depend on $\phi$ explicitly. As a result, the associated momentum is conserved:
\be \lab{Pi}
 \Pi=\frac{\partial {\cal{L}}}{\partial\theta'}= - \frac{\q' G_{tt} G_{\f\f}}{\cL} = constant.
\ee %
Furthermore, $\Pi$ depends only on $\q'$  and not $\q''$. Thus the equation of motion for $\q$ is first order. The general form of this equation is obtained from (\ref{Pi}) in terms of the metric components as,
\be\lab{thetaeq0}
\q^{'2} = \Pi^2 \frac{(-G_{tt} - \o^2 G_{\f\f} ) ( \rho^{'2} G_{\rho\rho} +G_{uu})}{G_{tt}G_{\f\f} (\Pi^2 + G_{tt} G_{\f\f})}.
\ee
Therefore the boundary condition $\q(\infty)=0$ is enough to determine the full solution (up to an overall sign which can be fixed by hand with no loss of generality).

On the other hand, a typical background metric does explicitly depend on
$\rho$, thus the equation of motion for $\rho$ will be second
order. One boundary condition is provided by eq. (\ref{bc}). The other will be
completely fixed by demanding regularity of the solution, just like in
\cite{Seattle, Gubser1} and \cite{Fadafan}, as follows: From (\ref{thetaeq0}) we see that the LHS is always positive definite. In order the RHS be positive definite as well, one needs to impose that the numerator and the denominator
changes sign at the same point. This usually happen at a finite $u=u_c$ because $G_{tt}$ in (\ref{thetaeq0}) is negative definite. One determines $u_c$ and the value of $\rho(u_c)  \equiv \rho_c$ at this point from the equations
\bea\lab{det1}
\Pi^2 + G_{tt}(u_c) G_{\f\f}(u_c) &=& 0,\\
G_{tt}(u_c) + \o^2 G_{\f\f}(u_c)  &=& 0 \lab{det2}
\eea
respectively. Consider any metric of the generic form
\be\lab{genmet}
ds^2 = b(u)\le( -dt^2 f_t(u) + d\r^2 + \r^2 d\f^2 + dx_i^2 + \cdots \ri).
\ee
Here the function $f_t(u) = 1$ in the low T (confined) phase of the dual QFT, and it is the blackness function of the black-hole in the high T (deconfined) phase.

The locus $u_c$ precisely corresponds to the point where velocity of the string  $\r(u)\o$ coincides  with the local speed of light observed
from infinity, $c(u) = \sqrt{f_t(u)}.$ This fact was already observed in \cite{Fadafan} and here we see that it directly generalizes to a large class of the string backgrounds.
This point also coincides with the location of a world-sheet horizon.  The latter is determined by substituting  (\ref{thetaeq0})  in (\ref{wsmet}) precisely as (\ref{det2}).
Again specifying to a metric of the general form (\ref{genmet}) this is given by
\be\lab{vc}
v_c \equiv \r(u_c) \o = f_t(u_c).
\ee
At this point it is important to note that {\em the world-sheet
horizon is present for rotating strings even in the low T phase when there is no black-hole horizon}.
This was already noted in a different set-up in \cite{MissingRef1} (see also \cite{MissingRef2} for a more recent discussion) and \cite{Das}.
In our case, it is given by the locus $\r(u_c) = 1/\o$. One can contrast this with the case of string in linear motion.
In that case, generally there would be no world-sheet horizon in the low-T phase, see e.g. \cite{GKMN4}.
In the dual field theory, the fact that there is a world-sheet horizon at low T means that there will be energy loss---in terms of Hawking radiation  from the world-sheet horizon\footnote{We refer the reader to \cite{Iancu}  and the references therein for a general discussion.}---even in the low T phase.


Quite generally, expansion of the equation of motion
\be\lab{rhoeq}
\frac{d}{du} \frac{\delta \cL}{\delta \rho'} = \frac{\delta \cL}{ \delta \rho},
\ee
around the point $u_c$ completely determines the first derivative $\rho'(u_c)$  in terms of $\o$ and $\Pi$.  Solving the equation from the point $u = u_c + \epsilon$ toward the boundary then fixes the solution completely\footnote{
More precisely, this determines the solution in the part $u>u_c$. The other section $u<u_c$ is also completely determined similarly by solving the equation starting from $u=u_c -\epsilon$.}. Finally, equating  the value $\rho(\infty) = L$ by the boundary condition (\ref{bc}) determines
$\Pi$ in terms of the parameters of the field theory problem $\o$ and $L$.

Energy loss rate is given by the world-sheet momentum :
\be\lab{el1}
\frac{dE}{dt}  = \Pi^\sigma _t = -\frac{\delta S }{\delta \6_{\sigma} X^0 } = - \frac{G_{tt} G_{\f\f} \o \q'}{2\pi \a' \cL },
\ee %
where, in the last equation, we used the general form of the Nambu-Goto action in terms of the metric functions. Using eq. (\ref{Pi}) above,
one then finds the simple general form,
\be\lab{el2}
\frac{dE}{dt}  = \frac{\Pi(\o,L) \o}{ 2\pi \a'}.
\ee %
We will be rather interested in the energy loss as a function of the field theory parameters $\o$ and $L$, which is determined by
the function $\Pi(\o,L)$. In general this can only be determined by numerical solution of the fluctuation equations as mentioned above.

There is another useful way of rewriting (\ref{el2}) in the high T phase. Using (\ref{det1}) and (\ref{det2}), equation (\ref{el2}) becomes,
\be\lab{el3}
\frac{dE}{dt}  = \frac{|G_{tt}(u_c)|}{ 2\pi \a'} = \frac{b(u_c) f_t(u_c)}{ 2\pi \a'}= \frac{b(u_c) v_c^2}{2\pi \a'},
\ee %
where we also specified to a general form (\ref{genmet}) and used (\ref{vc}). Now, for a generic form  with metric functions given by power-laws
\be\lab{genmet2}
b(u) = \le(\frac{u}{\ell}\ri)^{\g}, \qquad f(u) = 1-\le(\frac{u}{u_h}\ri)^\a,
\ee
one then finds,
\be\lab{el4}
\frac{dE}{dt}  = \frac{1}{ 2\pi \a'}\le(\frac{u_h}{\ell}\ri)^{\g}\frac{v_c^2}{(1-v_c^2)^{\frac{\g}{\a}}},
\ee %

\subsection{$D4/S^1$ system at finite temperature}
\lab{rev2}

The first example of a gravitational system that corresponds to a confining gauge theory under the gauge/gravity duality, was provided by Witten in his seminal paper \cite{Witten1}. The construction is based on N D4 branes wrapped on a circle of radius $R$. The fermionic fields on the brane acquire tree-level masses with choice of anti-periodic boundary conditions along the circle, and the scalar fields acquire masses at one-loop order. Therefore the supersymmetry of the original D4 brane system is completely broken by compactification on $S^1$.

The UV scale is given by the radius of the circle $R$, and as usual with the gauge/gravity correspondence the gravity approximation is valid when the 't Hooft coupling on the D4 brane theory is large, $\l_5\gg R$. In this limit the gravitational background is a solution to IIA string theory given by the metric, the  dilaton and the RR-four form fields as follows\cite{Witten1}:
\be\lab{lowT}
\begin{split} %
 ds^2&=(\frac{u}{\ell})^{3/2}(-dt^2+d\rho^2+\rho^2d\phi^2+dx_3^2+fdx_4^2)+(\frac{\ell}{u})^{3/2}
 (\frac{du^2}{f}+u^2d\Omega_4^2)\cr
 F_{(4)}&=\frac{2\pi N}{V_4}\epsilon_4,\ \ e^\phi=g_s(\frac{u}{\ell})^{3/4},\ \  \ell^3\equiv\pi g_sN_c\ell_s^3.
\end{split}\ee %
Here $V_4$ and $\epsilon_4$ are the volume of the unit for-sphere and the associated volume form, and,
\be\lab{lowf}
f_k=1-(\frac{u_k}{u})^3, \qquad \frac{u_k}{\ell} = \frac49 \frac{\ell^2}{R^2}.
\ee
The second relation follows from demanding absence of a conical singularity at the tip of the cigar $u_k$ that is spanned by $x_4$ and $u$. In these formulae above, $\ell$ is a typical length scale associated to the D4 brane geometry. We shall measure all dimensionful quantities in units of $\ell$ below.

The dual field theory on $D4/S^1$ is characterized by the UV cut-off scale $1/R$, the 3+1 dimensional 't Hooft coupling
$\l_4 = \l_5/2\pi R = 2\pi\ell_s g_s N/R$ and the confinement scale $u_k^{-1}$. The decoupling limit on the D4 branes is obtained by taking $\ell_s\ll 1$ as usual. In this limit the 10 D Newton's constant also vanish, hence the string interactions can be ignored as long as the dilaton is
not too large.  On the other hand $\l_4$  is given by
\be\lab{l5}
\l_4 = \frac{\l_5}{2\pi R}, \qquad with\qquad \frac{\l_5}{\ell}  = 4\pi \le( \frac{\ell}{\ell_s}\ri)^2.
\ee
In what follows we will keep $\ell \sim R \sim u_k \sim \cO(1)$ and $\ell_s \ll 1$. Thus the supergravity approximation is valid when $\l_5/\ell \sim \l_5/R \gg 1$.  Then, by (\ref{l5}) the SG limit means $\l_4\gg 1$, hence effective theory in 3+1 D should be strongly interacting. On the other hand, the confinement scale is around $1/R$, which is the same as the UV scale of the $D4/S^1$ theory.  It will be convenient for our purposes to define this scale---in units of $\ell$---as
\be\lab{mgap}
\Lambda_{QCD}  = \sqrt{\frac{u_k}{\ell}}.
\ee
This is a dimensionless parameter that parametrizes the mass gap in the theory.

In the SG limit the 3+1 D field theory involves infinitely many KK modes that cannot be decoupled from the gauge modes. In the opposite limit $\l_5/R \ll 1$ however, the confinement scale is exponentially smaller than the KK scale and the theory flows to pure $SU(N)$ Yang-Mills in 3+1 D at energies smaller than $1/R$\footnote{One can easily see this from the beta-function equation in the gauge theory. One loop approximate solution in 3+1 D yields $M_{glue} \sim R^{-1} \exp( 2\pi R/\l_5)$ where the UV scale is set as $1/R$. Clearly, for a parametric separation between the glue and the KK scales, one needs to go outside the validity of the SG approximation.}. The stringy corrections cannot be ignored in this limit. However, It is believed that the theory in these two opposite limits are continuously connected, and in this sense the $D4/S^1$ set-up even in the SG approximation corresponds to a theory in the same universality class as the pure non-supersymmetric 3+1 Yang-Mills.

Another concern is that there is an upper bound on the variable $u$ because from (\ref{lowT}) requirement of a small dilaton means
$u/\ell \ll g_s^{-4/3}$. This however is always satisfied in the large N limit, provided that we first take $N\to\infty$. One can easily see this
by noting from the equations above that, requirement of small dilaton is
\be\lab{sdil}
\frac{u}{\ell} \ll  \pi^{\frac43} \le( \l_4 \frac{R}{2\ell} \ri)^{-2} N^{\frac43}.
\ee
Although the quantity in the brackets is large in the SG limit, it is kept finite, and $N$ is taken to $\infty$ first. Thus the upper limit on $u$ will play no role in what follows.

Temperature in the field theory is introduced by compactifying the Euclidean time on a circle with periodicity $1/T$: $t_E \sim t_E + 1/T$, where $t_E = it$ is the Euclidean time. Thus, the theory for sufficiently low temperatures is dual to the geometry given by a cylinder in the $(u, t_E)$ plane and a cigar in the $(u, x_4)$ plane. There exists another geometry with the same near boundary asymptotics, that is obtained by exchanging the $t_E$ and the $x_4$ coordinates \footnote{The four-form field and the dilaton are exactly the same as in (\ref{lowT}).}. This is the black-hole geometry with a horizon located at $u_h$:
\be\lab{highT}
 ds^2=(\frac{u}{\ell})^{3/2}(-fdt^2+d\rho^2+\rho^2d\phi^2+dx_3^2+dx_4^2)+(\frac{\ell}{u})^{3/2}
 (\frac{du^2}{f}+u^2d\Omega_4^2)
\ee %
where $\ell$ is given as in (\ref{lowT}). The blackness function is
\be\lab{highf}
f_h=1-(\frac{u_h}{u})^3, \qquad \frac{u_h}{\ell} = \le(\frac{4\pi}{3}\ri)^2 (T\ell)^2.
\ee
The latter expression above is again determined by demanding absence of a conical singularity at the horizon $u_h$.
The free energies associated with the (Euclidean versions of)solutions (\ref{lowT}) and (\ref{highT}) are obtained by the IIA supergravity action evaluated on these geometries. Clearly the difference between the actions vanish when the perimeters of the $x_4$ and the time  circles become identical. This corresponds to a (first order) confinement-deconfinement transition\footnote{Reference \cite{Takeshi} questions the validity of the identification of this black-hole phase with the deconfined phase of the $D4/S^1$ gauge theory and concludes that, in fact another geometry i.e. the localized $D3$ brane geometry corresponds to the true deconfined phase of the gauge theory. The arguments  in this paper seem to be convincing and it would be very interesting to study the energy loss in this new background as well. However, our purpose in this work is to discuss the qualitative features of the low and high T phases in a non-conformal background, and the black-hole solution below certainly serves for this purpose. We offer more comments about the high T phase in the beginning of section \ref{rothigh}.}  in the dual gauge theory at:
\be\lab{Tc}
T_c = \frac{1}{2\pi R}.
\ee
The theory for $T<T_c$ is confined and described by the geometry (\ref{lowT}), whereas, for $T>T_c$ the thermodynamics ensemble is dominated by (\ref{highT}).
One useful dimensionless combination of parameters is given by the ratio $u_h/u_k$
that, in terms of physical parameters reads as,
\be\lab{uhuk}
\frac{u_h}{u_k} = \le(\frac{T}{T_c}\ri)^2.
\ee
\begin{figure}[ht]
\includegraphics[width=3 in]{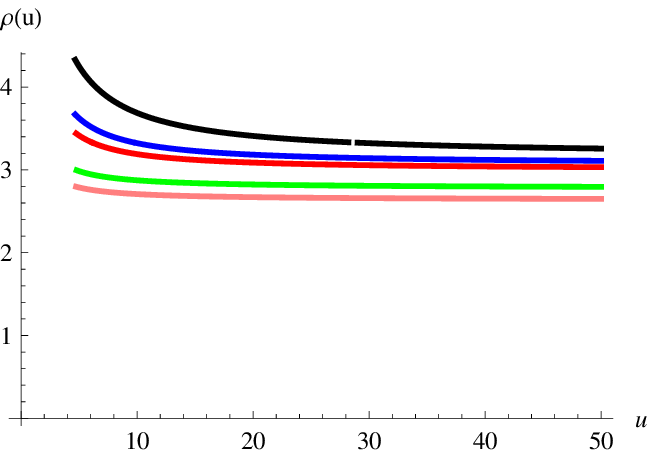}%
\includegraphics[width=3 in]{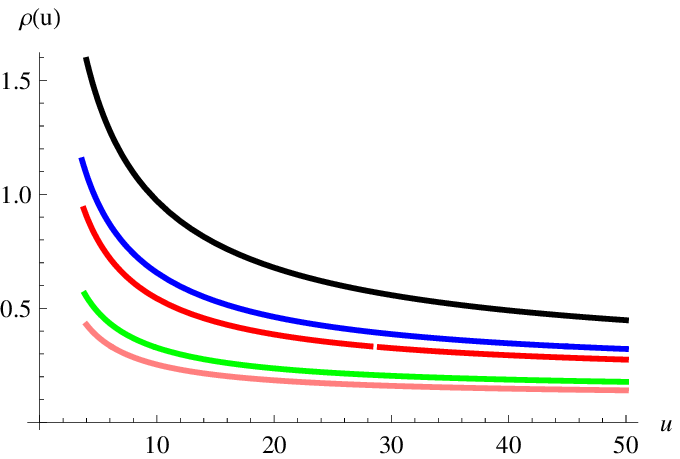}%
\caption{The local radius of the rotating string in the low T phase as
  a function of the radial coordinate $u$. The different curves in
  each figure correspond to different choices of the world-sheet
  momentum $\Pi=500,100,50,10,5$ (from above to
below). Left: $\o/\Lambda_c = 0.3$ Right: $\o/\Lambda_c = 3$.  }
\lab{lowTrho}
\end{figure}%

Flavor sector in the theory described above is represented by flavor
$D8-\bar{D8}$ branes that span the same field theory directions $x^i$,
$i=0,\cdots 3$ and localized at two arbitrary points that can be
chosen as $x^4=0$ ($D8$) and $x^4=r$ ($\ov{D8}$) on the $S^1$ in the
background \cite{SS}. A hard quark(anti-quark) probe traveling through
the gluon plasma is then represented in this set-up as a probe string
attached to the flavor $D8$ ($\overline{D8}$) branes. We will not deal
with the dynamics of flavor sector in this paper. Therefore the number
of flavor D8 branes is much smaller than the number of color D4 branes
corresponding to the quenched approximation in the dual field theory,
where the quarks do not propagate in loops and treated as external
probes\footnote{Study of QGP beyond the quenched limit in the
  holographic setting is quite interesting \cite{Carlos}. See \cite{Bigazzi} for a review.}.

\section{Rotating quarks at low temperature}
\lab{rotlow}
\subsection{Rotating string solution}

We first consider energy loss of a rotating probe in the $D4/S^1$ theory in the low T phase, i.e. when $T<T_c$. The theory is in the confined phase, therefore one expects energy loss only associated with radiation due to acceleration in circular motion of the probe.
This is because the color charge in the low T states of the theory is always confined inside hadrons, hence the flux lines emanating from the charged probe cannot end on anywhere in the medium, thus one does not expect any drag associated with interactions with the medium. A similar fact in the case of linear constant motion of a quark was observed in \cite{Talavera}.

\begin{figure}[ht]
\lab{shapelow}
\centerline{\includegraphics[width=2.7 in]{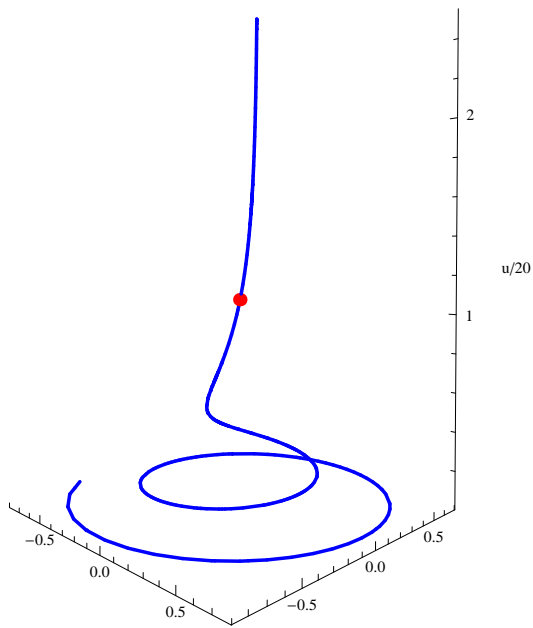}}%
\caption{Shape of the rotating string solution in the low T phase, for
  the choice of parameters $\frac{\omega}{\Lambda_c}=8$, $\Pi=12$ (in
   units $\ell=1$.) The red dot represents the location $u_c$ that
   corresponds to the world-sheet horizon.}
\end{figure}%

The dual gravity theory should be able to reproduce this feature for consistency of the gauge-gravity correspondence, as we confirm in this section. We begin by constructing the rotating string solution in the low T background (\ref{lowT}). The on shell action follows from substituting (\ref{lowT}) in (\ref{NG}) as
\be\begin{split} \lab{lowNG}
 S&=\frac{1}{2\pi\alpha'}\int d\tau d\sigma\sqrt{(\frac{u}{\ell})^3\rho^2\theta'^2+(\frac{u}{\ell})^3(1-\rho^2w^2)(\rho'^2+\frac{1}{(\frac{u}{\ell})^3f})}\cr
 &\equiv\frac{1}{2\pi\alpha'}\int d\tau d\sigma{\cal{L}}.
\end{split}\ee %
In what follows we present our computations in terms of dimensionless variables. This is trivially done by omitting  all factors of $\ell$ in the equations. By an abuse of notation we will denote the dimensionless variables also by same symbols. At the end of the computation, one can
reinstate dimensions simply by the substitutions:
\be\lab{subs}
u\to \frac{u}{\ell},\,\, \rho \to \frac{\rho}{\ell}, \,\, \o \to \o \ell, \,\, \Pi \to \frac{\Pi}{\ell}.
\ee
Then the equation of motion for $\theta$ (\ref{Pi}) gives
\be \lab{thetaeq1}
 \Pi=\frac{u^3\rho^2\theta'}{\ell^3\cal{L}}.
\ee %
Upon use of (\ref{lowNG}), or directly from (\ref{thetaeq0}) one obtains
\be \lab{thetaeq2}
 \theta'^2=\frac{\Pi^2(1-\rho^2w^2)(\rho^{'2}+\frac{1}{(\frac{u}{\ell})^3f})}{\rho^2(\rho^2(\frac{u}{\ell})^3-\Pi^2)}.
\ee %
Given the function $\rho(u)$ and the boundary condition for $\q$ in (\ref{bc}) this equation completely determines $\q$.  The equation of motion
for $\rho$ on the other hand is given by
\be\begin{split} %
 \rho''&-\frac{\rho'(3\rho^2u_k^3+2u^4f\rho\rho'+3\Pi^2-6u^3\rho^2-3u^6f^2\rho^2\rho'^2)+2u\rho}{2uf(\rho^2u^3-\Pi^2)}\cr
 &-\frac{1+u^3f\rho'^2}{u^3f\rho(1-\rho^2\o^2)}=0
\end{split}\lab{rhoeq1} \ee %
The regularity condition of (\ref{det1}) and (\ref{det2}) yields
\be\lab{uc}
\rho(u_c) = \rho_c = \frac{1}{\o} \qquad at\,\,\, u_c=\ell(\Pi
\o)^{2/3}.
\ee
In order to find the shape of the string solution,
one has to solve the differential equation (\ref{rhoeq1})
numerically. This is done by starting at $u=u_c +\eps$
for an $\eps\ll1$ and solving the equation towards the boundary at
$u\to\infty$ (in practice it suffices to solve until a  large enough
value of $u$ such that $\rho(u)$ ceases to change further as $u$ is
varied).  Substituting $u=u_c +\eps$ in (\ref{rhoeq1}) and expanding
in $\eps$, the leading term determines $\rho'$  as a
solution of the following algebraic equation
\footnote{Generically, there is a single acceptable solution with real and negative $\r'$. Negativity requirement comes from the fact that $r$ is supposed to decrease towards the boundary.}:
\be %
 \big(\rho'^2(u_k^3-\Pi^2 \omega^2)-1\big)\Big(4
 \rho'\Pi\omega^2+3(\Pi\omega)^{1/3}\big[1+\rho'^2(u_k^3-\Pi^2 \omega^2)\big]\Big)=0 %
\ee
Therefore, once $\Pi$ and $\o$ are given, the entire solution $\rho(u)$ is determined completely\footnote{In the other range $u_k<u<u_c$ the solution is determined similarly by expanding in $u=u_c -\eps$.}.
We plot the function $\r(u)$ for various different choices of $\Pi$ and $\o$ in figure 1. These solutions are similar to the conformal case in \cite{Fadafan}: The local radius of the rotating string decreases as one approaches the boundary, and it increases as the parameter $\Pi$ is increased.
Once this is found, the angular dependence is determined by solving (\ref{thetaeq2})
with the boundary condition (\ref{bc}). A sample solution is presented
in figure 2 where we also display the location of the world-sheet
horizon by the red dot.

\begin{figure}[ht]
\lab{figlmin}
\begin{center}
\includegraphics[width=2.8in]{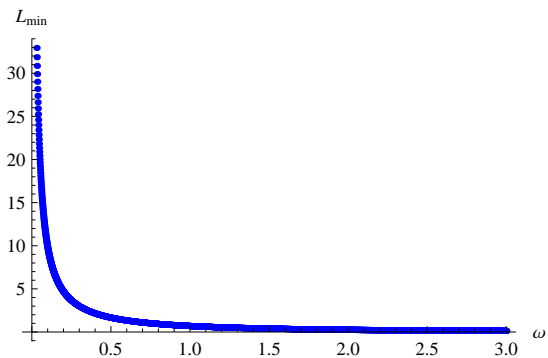}%
\end{center}
\caption{Minimum allowed value of the radius of rotation as a function
  of frequency (in units $\ell=1$). In the physical units $\frac{\omega}{\Lambda_c}$ the range of
  frequency corresponds to from 0.013 to 1.34.}
\end{figure}

One important point is that, one has to choose the range of $\Pi$ and $\o$ such that $u_c$ is always  bigger than $u_k$. From equation
(\ref{uc}), we learn that this means
\be\lab{restrict}
\Pi \o > \le(\frac{u_k}{\ell} \ri)^{\frac32}.
\ee
Given $\o$ this provides a lower bound on the allowed values of
$\Pi$, vice versa. On the other hand we numerically observe in figure 3 that the
length $L$ is a monotonically increasing function of $\Pi$ at fixed
$\o$. Therefore, a lower bound on $\Pi$ implies a lower bound on
$L$. We call this value $L_{min}$ and plot is as a function of $\o$ in
figure 3.
\begin{figure}[ht]
\begin{center}
\includegraphics[width=3 in]{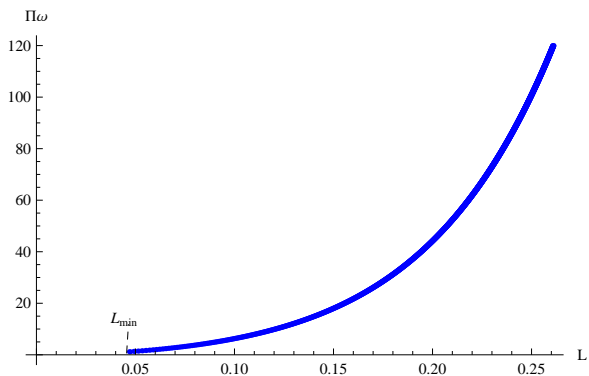}%
\caption{Energy loss rate $\Pi\o = 2\pi \a' dE/dt$ in the low T phase
  as a function of the radius of rotation for a large choice of frequency
  $\frac{\omega}{\Lambda_c}=3$. We denote the minimum allowed
  value of the radius as $L_{min}$. }
\end{center}
\end{figure}%

\subsection{Energy loss}
\lab{subel1}

Once the string solution is found, the energy loss of the rotating quark in the dual field theory is given by (\ref{el2})
\be \lab{ellow2}
 {2\pi\alpha'}\frac{dE}{dt}=\o\Pi(\o,L)=(\frac{u_c}{\ell})^{3/2},
\ee %
where we also used (\ref{uc}) in the last equality.

We are, however, interested in $dE/dt$ as a function of $\o$ and $L$
rather than $\Pi$. One needs to numerically evaluate the function
$\Pi(\o,L)$.
 This is achieved by solving (\ref{rhoeq1}) as described at the end of
 the previous section for a given $\Pi$ and $\o$ and read off the
 value $L$ from the asymptotics of $\rho(u)$ for  large enough $u$.

The result of this numerical calculation is shown in figures 4 and
5. In figure 4 we plot the energy loss rate as a function of the
radius of rotation $L$.  It is clearly a monotonically increasing
function of $L$.

Figure 5 presents the same function at different
values of the frequency. As clearly seen from this figure changing
$\o$ only changes the allowed range of $L$. The allowed range
increases ($L_{min}$ decreases) with increasing $\o$. This is because,
from equation (\ref{restrict}) the minimum value of $\Pi$ decreases,
therefore $L_{min}$ also decreases by monotonicity (figure4).
The form of the curve stays the same for different $\o$ is
becuase the only dimensionally meaningful quantity is the ratio
$\o/\Lambda_c$, therefore increasing $\o$ is equivalent to decreasing
 $\Lambda_c$ (or equivalently, increasing $u_k$). The latter change can be absorbed into a redefinition of
 the $u$ variable, hence does not change the form of the string
 equation of motion but only the allowed range of $u>u_c$.

\begin{figure}[ht]
\begin{center}
\includegraphics[width=4 in]{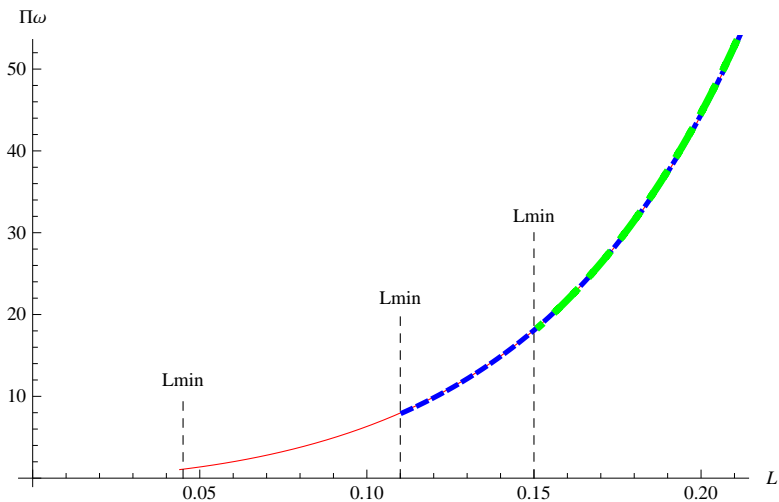}
\caption{The energy loss rate in the low T phase for small choices
  of frequencies $\frac{\omega}{\Lambda_c} =1$, 0.85 and 0.6 for the blue (solid),
  red (dashed) and the green (dot-dashed) curves.}
\end{center}
\end{figure}%

\subsection{Energy loss at low $T$: discussion}
\lab{lowrad}

Motivated by \cite{Fadafan}, we ask how much of the energy loss is due to linear drag and how much is due to radiation\footnote{In \cite{Fadafan} the question is asked is asked in case of the ${\cal N}=4$YM theory which is qualitatively similar to the high T phase of the $D4/S^1$ plasma which we discuss in the next section.}.
As discussed at the beginning of this section, we do not expect any
linear drag component in the low energy phase, and the entire energy
loss should be due to radiation from the rotating probe.
We present four arguments to support this view based on calculations on the gravity side.

\bz

\item A simple argument can be made by taking the ``linear drag limit'' of the rotating string solution, as in \cite{Fadafan}, that is, $\o \to 0$, $L\to \infty$, $v = \o L = constant$.  In this limit one should use the rescaled variable $v(u) = \o \r(u)$ instead of $\r(u)$ which diverges. The leading term in the equation  of motion for $\r$ (\ref{rhoeq1}) then becomes
$v'(u)= 0$. Therefore it should be given by its value at $u_c$ that is $v(u) = \o \r_c$. On the other hand, by (\ref{uc}) this means $v=1$, that contradicts our assumption that the end point of the string describes a probe moving at some velocity  $v<1$. Therefore, there is no smooth linear drag limit in the low T phase.

\item Related to the argument above, one notes the following. As mentioned in section \ref{rev1}, there exists
a horizon on the world-sheet of a rotating string solution. This horizon is given by $\r(u_c) = 1/\o$ in the low T phase. Quite generally AdS/CFT associates Hawking radiation from a world-sheet horizon to energy loss of the probe on the boundary theory. On the other hand one can show that the presence of this horizon is totally due to rotation and not linear drag. This is because we know that the horizon disappears in the linear drag limit discussed above. In support of this view, one can also note that the horizon disappears in the limit $\o\to 0$ because $\r(u_c)= 1/\o\to\infty$ cannot happen for any $\infty<u_c<u_k$. This is in accord with the fact that syncrotron radiation disappears in the limit $\o\to 0$.


\item In the next section we show that, the limit $\o\to\infty$ of the energy loss in the high T phase is entirely accounted for, by the energy loss in the low T background. As, one expects that the energy loss in the high T phase should be dominated by radiation in the large $\o$ limit, this provides another argument in support of our view that low T energy loss is completely due to radiation, see equation (\ref{UVlim}).

\item Perhaps the strongest argument can be made as follows: We recall the lower limit (\ref{restrict}) on the energy
  loss rate of the quark:
\be\lab{lowlim}
2\pi\a' \frac{dE}{dt} = \Pi \o > \Lambda_{QCD}^3,
\ee
where $\Lambda_{QCD}$ is a dimensionless parameter that characterizes
the mass gap in the confined theory and defined in (\ref{mgap}).
This lower bound is in complete agreement with our expectations on the dual field
theory side. In the low T phase the theory is confined. On the other
hand when a probe accelerates, it should radiate its energy in gluon
quanta. However the gluons cannot have arbitrarily small energy
because of confinement: they should turn into jets of
glueballs. Indeed what we observe from eq. (\ref{lowlim}) is that,
precisely the lower limit on the energy loss is given by the mass gap
of the confining theory (\ref{mgap})! Consistency of this picture
also requires that energy loss can only be in terms of radiation in the low
T phase: if it was also by drag, then there would be no reason for a lower
bound because the drag force can be made arbitrarily small by making the
velocity of the quark $v = L \o$ arbitrarily small. Therefore, we see
that the only possible interpretation of a lower bound eq. (\ref{lowlim}) in the
energy loss is if it is totally due to radiation.

\ez

We conclude that the rotating string solution in the low T phase
should directly measures the energy loss of a rotating probe in the
vacuum theory due to radiation. The latter can be calculated in
perturbative QFT. It is given by the Lienard potential of a charged
rotating particle\cite{Lienard} in case of electromagnetic
radiation. An immediate consequence of our result is that, the
gauge-gravity correspondence provides a way to calculate the Lienard
potential for the syncrotron radiation in a complicated field theory
given by the $D4/S^1$ branes at strong coupling! The result is given
in figures 4 and 5. Our result is only numerical and it is beyond our technical ability to determine the dependence of the Lienard potential in $\o$ and $L$ analytically.

The only known analytic result at strong coupling is in the case of ${\cal N}=4$ YM theory, that is obtained by the dual rotating string calculation in the vacuum AdS solution \cite{Mikhailov} (specified to rotation with constant frequency and radius):
\be\lab{radan}
\frac{dE}{dt}\bigg|_{{\cal N}=4, radiation}  = \frac{\sqrt{\l}}{2\pi} \frac{v^2\o^2}{(1-v^2)^2}.
\ee
This turns out to be just equivalent to Lienard's result for electromagnetic syncrotron radiation, except for a simple replacement of the proportionality constant.
\begin{figure}[ht]
\begin{center}
\lab{Compar}
\includegraphics[width=2.5in]{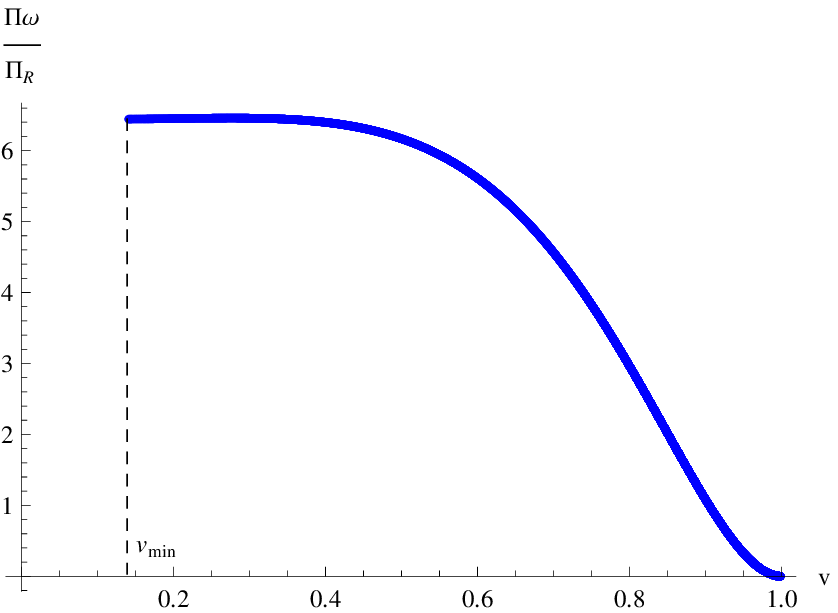}
\hspace*{.5cm}
\includegraphics[width=2.5in]{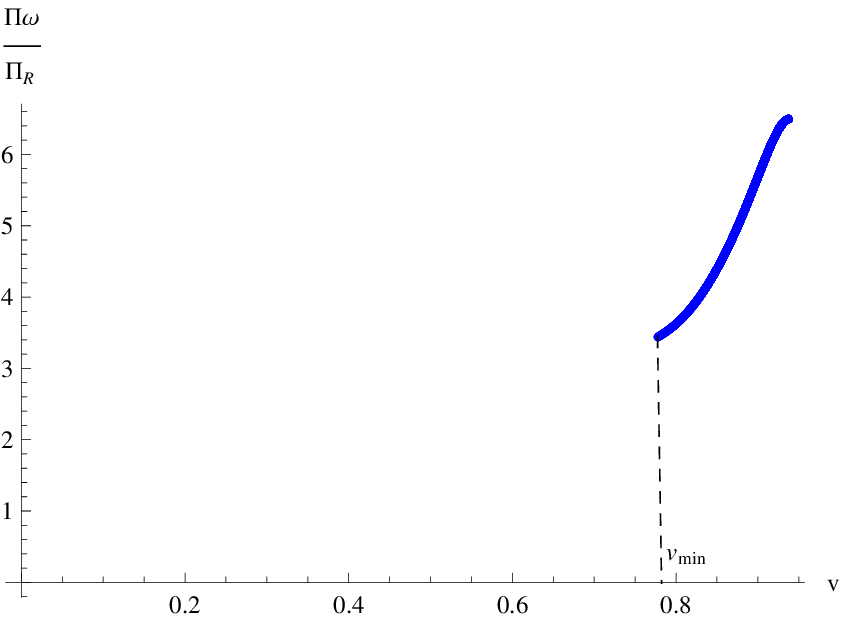}\\%
\end{center}
\caption{The numerical proof that a simple Lienard type formula
  (\ref{piar}) for syncrotron radiation does not hold for the $D4/S^1$
  plasma. The ratio of the energy loss in the low T phase versus
  (\ref{piar}) is plotted  for two different frequencies
  $\frac{\omega}{\Lambda_c}=3$ (left) and 0.3 (right) as a function of
  the velocity $v = \o L$. The minimum
  allowed value of the velocity $v_{min} = \o L_{min}$ is also displayed.}
\end{figure}%

It is tempting to ask whether the form in (\ref{radan}) is also attained in a more complicated theory such as the $D4/S^1$ theory. For that purpose we define the function
\be\lab{piar}
\Pi_R = \frac{\sqrt{\l}}{2\pi} \frac{v^2\o^2}{(1-v^2)^2},
\ee
and we plot in figure 6 the ratio of our result for the energy loss in
the vacuum $D4/S^1$ theory and the RHS of equation (\ref{radan}) where
we set $\sqrt{\l} = 2\pi$ for simplicity. We observe that the ratio
depends on v and $\o$, therefore a simple analytic formula as in
(\ref{radan}) does not work in the confining case\footnote{This
  conclusion of course relies on our arguments above that the low T
  energy loss is totally due syncrotron radiation.},
neither for small nor for large frequency.
\begin{figure}[ht]
\begin{center}
\includegraphics[width=2.7in]{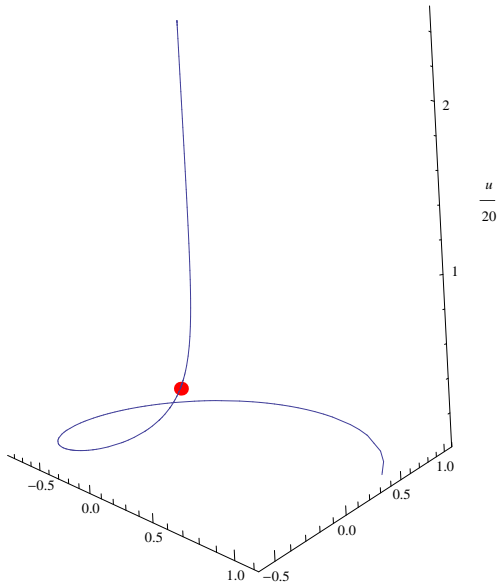}%
\end{center}
\caption{Shape of the rotating string solution in the high T phase for
  a choice of large frequency $\frac{\o}{T} = 15$. The plot is for the
  choice of $\Pi=12$ (in units of $\ell=1$). The red dot represents
  the location $u_c$ that corresponds to the world-sheet horizon.}
\end{figure}%

\section{Rotating quarks at  high temperature}
\lab{rothigh}
In the high T phase $T>T_c$ the field theory is in the deconfined phase. As in the case of ${\cal N}=4$ plasma
one expects two components in the rate of energy loss \cite{Fadafan}: one due to syncrotron radiation and one due to drag in the deconfined medium. Indeed, our results in this section will be qualitatively similar to \cite{Fadafan}.

Before going into details of this investigation let us discuss certain properties of this phase\footnote{We thank Takeshi Morita for a discussion on this point.}  in view of ref. \cite{Takeshi}. This paper suggests that the ``high T deconfined" phase of the 
$D4/S^1$ field theory is in fact a different geometry based on a localized $D3$ brane background. The reason for this identification is  as follows. Ref. \cite{Takeshi} argues that in order the KK reduction on $S^1$ to make sense one needs the Wilson 
loop around this circle to have non-trivial expectation value. This is indeed the case in the low T phase we discussed above, because the submanifold of (\ref{lowT}) spanned by the $x^4$ and $u$ directions form a cigar, hence a Nambu-Goto string wrapping this cigar (appropriately renormalized in the UV) has finite action, corresponding to a finite expectation value for the dual Wilson loop operator \cite{Witten1}. This is not the case in the black-hole solution (\ref{highT}), thus the Wilson loop around the $x_4$ circle vanishes. This means that the high T phase (\ref{highT}) is intrinsically a 5D theory. Ref. \cite{Takeshi} further argues that the true deconfined phase of the $D4/S^1$ field theory should be a localized D3 brane geometry in the T-dual IIB picture. 

Although the arguments in \cite{Takeshi} are convincing, we shall still perform our investigation of the high T phase in the background given by (\ref{highT}). The reason is two-fold. Firstly, we are interested in studying the deconfined phase of a non-conformal field theory at strong coupling, and (\ref{highT}) certainly describes a deconfined phase\footnote{This is because the Wilson loop around the Euclidean time direction is finite, hence the dual Polyakov loop has non-vanishing VeV.}. Thus it is suitable for our qualitative investigation here. Secondly, it is highly non-trivial to study the energy loss in the suggested localized D3 geometry, because the geometry depends on Euclidean time. Therefore, for the purpose of this paper the ``deconfined phase" corresponds to the background (\ref{highT}).            

\subsection{The rotating string}
\lab{subel3}
We begin by describing the rotating string solution in the high T background (\ref{highT}).
The on shell Nambu-Goto lagrangian is obtained by substituting (\ref{highT}) in (\ref{NG}) as
\be \lab{highNG}
 {\cal{L}}=\sqrt{(\frac{u}{\ell})^3\rho^2\theta'^2f_h+(\frac{u}{\ell})^3(f_h-\rho^2\o^2)(\rho'^2+(\frac{u}{\ell})^{-3} f_h^{-1})},
\ee %
where the blackness function $f_h$ is given by (\ref{highf}) with the horizon location $u_h$.
As in the previous section we shall omit all factors of $\ell$ in the
equations. They can be trivially reinstated as in eq.(\ref{subs}) \footnote{We will denote the dimensionless variables also by same symbols as the original dimensionful ones with a slight abuse of notation.}.
\begin{figure}[ht]
\lab{highTrho}
\begin{center}
\includegraphics[width=2.8in]{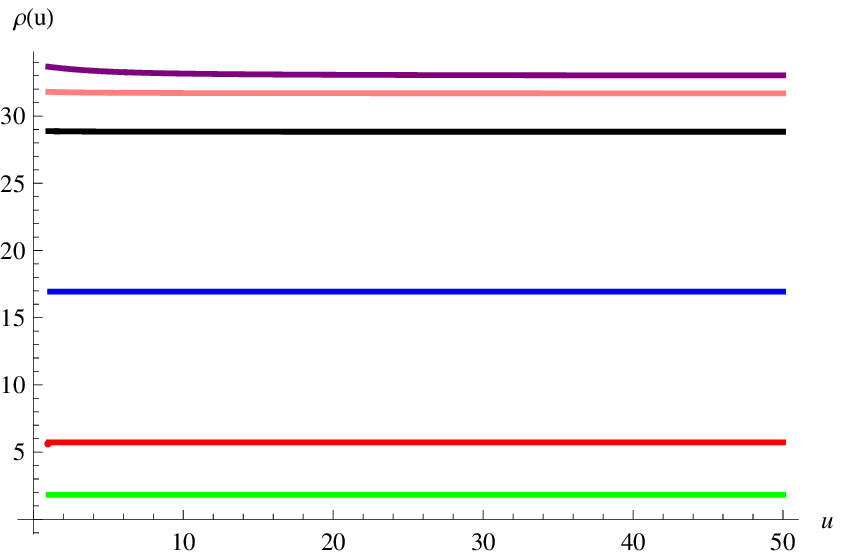}%
%
\includegraphics[width=2.8in]{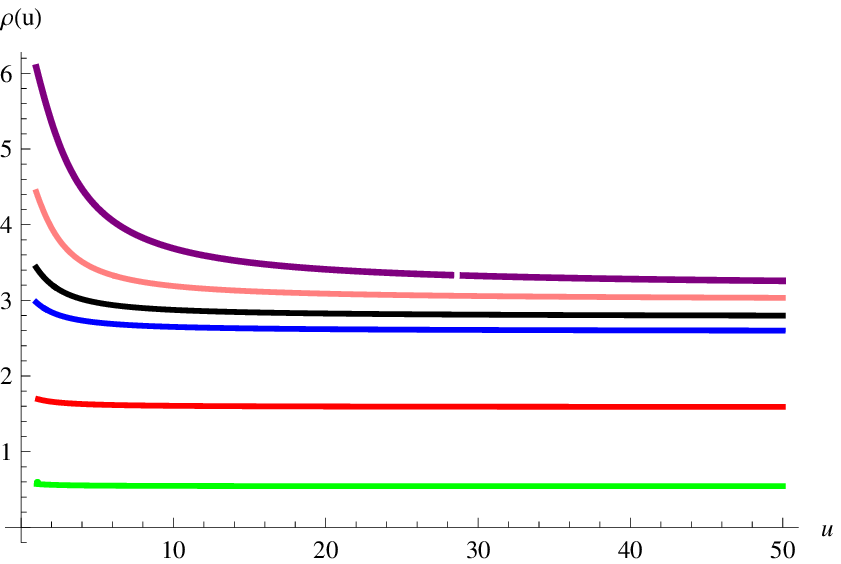}\\%
%
\includegraphics[width=2.8in]{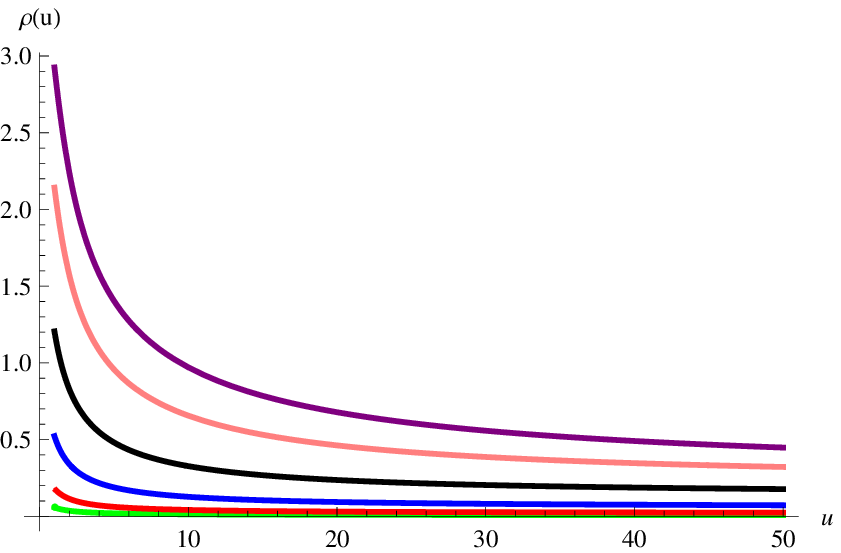}%
\includegraphics[width=2.8in]{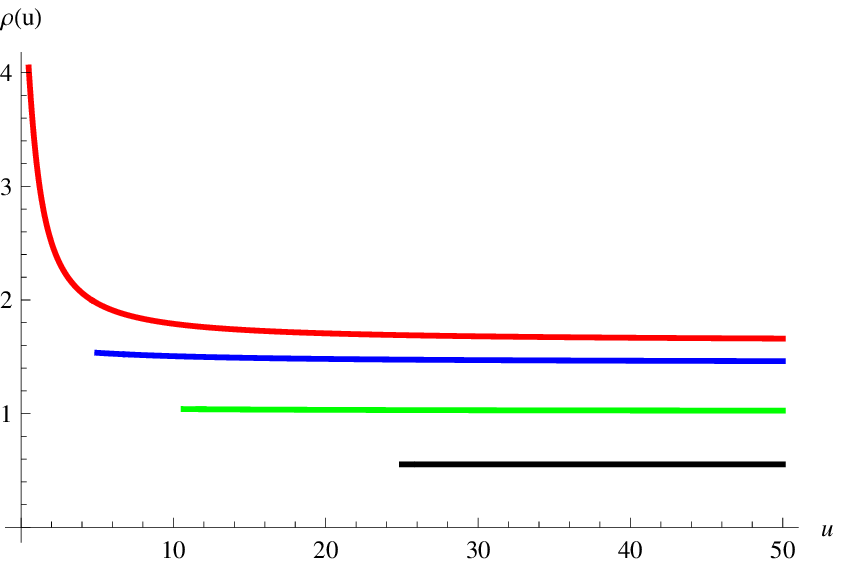}%
\end{center}
\caption{Local radius of the rotating string as a function of the
  radial coordinate for various different choices of parameters:
  $\Pi=500,100,50,10,1,0.1$ (from above  to below) in all of the
  figures except the bottom right, and $\frac{\o}{T} =
  0.13$ (top left), $\frac{\o}{T} =
  1.3$ (top right), $\frac{\o}{T} =
  13$ (bottom right). Bottom right figure is a comparison of different
   choice of $\frac{\omega}{T}=2, 0.93, 0,75, 0.42$ (from above to
   below), for a fixed angular momentum $\Pi = 20$ in units of $\ell=1$.}
\end{figure}%

The equation of motion for $\theta$ follows from (\ref{thetaeq0}) as
\be \lab{thetaeq3}
 \theta'^2=\frac{\Pi^2(f_h-\rho^2\o^2)(\rho^{'2}+u^{-3} f_h^{-1})}{\rho^2 f_h \big(\rho^2 u^3 f_h-\Pi^2\big)}
\ee %
The equation of motion for the radial variable $\r(u)$ is given by
\bea
&&\rho''+ \frac{2u^4 \rho^2 + 3\rho u^3(\Pi^2 - u^3 f \rho^2 - u^3\rho^2)\rho' + 2u^7\rho^2 f \rho^{'2} - 3u^9\rho^3f \rho^{'3}}{2u^3\rho(\Pi^2 - u^3 f \rho^2)} \\ {}&&+ \frac{2u+3u_h^3 \rho \rho' + 2u^4 f \rho^{'2} +3u^3 u_h^3 \omega^2 \rho^3 \rho^{'3}}{2u^3 \rho (f-\omega^2 \rho^2)}  =0. \lab{rhoeq2}
\eea
Just as the low T solution above, there is a special point $u_c$ where the denominator of the RHS of (\ref{thetaeq3}) changes sign. In order
the LHS of this equation to be positive definite then, the value of
$\rho_c\equiv \rho(u_c)$ should be chosen accordingly as in the low T
phase above. One finds,
\bea\lab{uc2}
 u_c&=&\bigg(\frac{\Pi \o}{2}+\frac{1}{2}\sqrt{4 u_h^3+\Pi^2\o^2}\ \bigg)^{\frac23} \\
\lab{rc2}  \rho_c&=&\frac{1}{\o}\sqrt{\frac{2\Pi \o}{\Pi w+\sqrt{4u_h^3+\Pi^2\o^2}}}
\eea %
We note that these equations have exactly the same form as those of
\cite{Fadafan} if one defines a dimensionless frequency ${\cal \o} =
\o/u_h^\frac32$ analogously.


The numerical evaluation of $\rho(u)$ and $\q(u)$ is completely
analogous to the previous section. We note that, unlike in the low T phase, here there is no restriction
on the minimum value of the energy loss $\Pi\o$ of the quark.
Such a restriction would come from the requirement that $u_c>u_h$
but from eq. (\ref{uc2}) we see that for any positive value of $\Pi\o$
this condition is satisfied automatically.

A sample shape of the string is given in figure 7.
Profile of the radial function $\rho(u)$ of the string, as shown in
figure 8 exhibits a similar behavior as in the low T phase, figure 1.

\subsection{Energy loss}
\lab{subel2}

The rate of energy loss of the rotating quark in the high T plasma is determined by the world-sheet momentum of the string solution found above. This is given by the general formula (\ref{el4}) which for the geometry (\ref{highT}) becomes
\be \lab{el41}
 {2\pi\alpha'}\frac{dE}{dt}=(\frac{u_h}{\ell})^\frac32\frac{v_c^2}{\sqrt{1-v_c^2}}.
 \ee %
Here we reinstated the units $\ell$ and used the equation (\ref{uc2}). $v_c$ is the local velocity of the string at the world-sheet horizon
$u_c$, i.e., $v_c = \r_c \o$ with $\r_c$ given by (\ref{rc2}). It determines the dependence of the energy-loss on $L$ and $\o$.
Using the definitions in section \ref{rev2}, the ``dimensionless'' energy-loss rate can be written in {\em physical parameters}  as,
\be \lab{el5}
\ell^2 \frac{dE}{dt}= \frac{32 \pi^2 \l_4}{9} \frac{(T\ell)^3}{\Lambda_{QCD}} \frac{v_c^2}{\sqrt{1-v_c^2}}.
 \ee %
Here $\l_4$ is the 't Hooft coupling in the dual 3+1 D field theory (\ref{l5}) and $\Lambda_{QCD}$ is the dimensionless mass-gap parameter defined in (\ref{mgap}).
We present our numerical findings for the various temperatures and frequency in figure 9.

\begin{figure}[ht]
\lab{PiDifHigh}
\begin{center}
\includegraphics[width=3in]{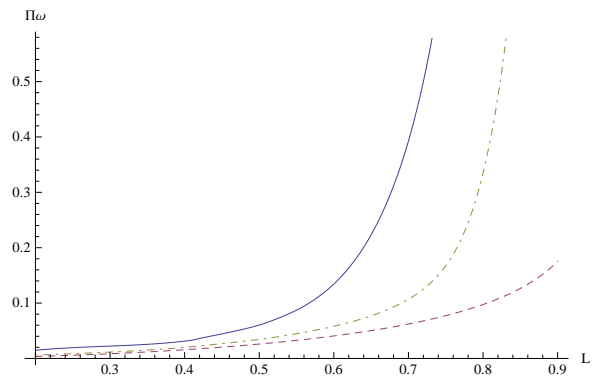}%
\caption{Energy loss rate $\Pi\o = 2\pi \a' dE/dt$ in the high T phase
  as a function of the radius of rotation for the choices of different
  $\o/T$=0.3 , 2, 3 (from below to above).}
\end{center}
\end{figure}%

As noted below eq. (\ref{rc2}) there is no lower bound on the rate of
energy loss of the quark, unlike in the low T phase, see eq. (\ref{lowlim}). This is in perfect
agreement with what we expect from the field theory dual. In the high
T phase the theory is deconfined and an accelerating quark loses energy in possible
ways:  by radiation due to acceleration and by the drag force that
arises from interactions with the deconfined medium surrounding the
quark. The latter can be arbitrarily small as the velocity of the
quark can be chosen arbitrarily small. Similarly, radiation due to
acceleration can also be arbitrarily small because the theory is deconfined and the quark can
radiate energy by gluons that can have arbitrarily small energies,
unlike in the confined case. One can distinguish these two contributions in the two separate limits. This is what we
discuss next.

\subsubsection{The IR and the UV limits}

As in section \ref{lowrad} for the low T phase, it is interesting to study the energy loss rate in the various limits.  Here we focus on the
two limits, that were already studied in \cite{Fadafan} in the case of the conformal ${\cal N}=4$ plasma. These limits are:
\ben
\item The IR or the ``liner drag dominated limit'': $\o\to 0$, $L\to\infty$, $v$ kept constant.
\item The UV or the ``radiation dominated limit'': $\o\to \infty$, $L\to 0$, $v$ kept constant.
\een
\begin{figure}[ht]\lab{Khoob}
\begin{center}
\includegraphics[width=3 in]{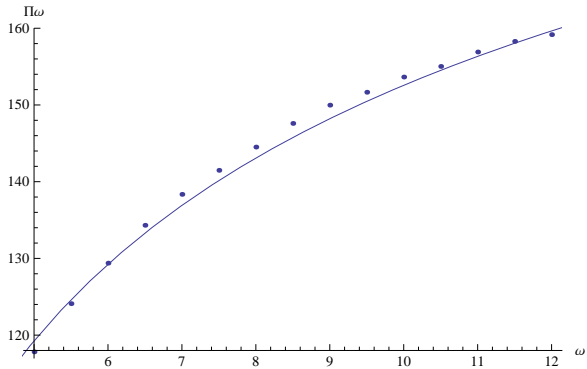}%
\caption{Energy loss $\Pi \o$ as a function of $\o$ in the limit $v=
  \o L$ is kept constant. The fit function reads $93\log\log\o + 75$.}
\end{center}
\end{figure}%
In the following we analyze both limits in detail.
\paragraph{Linear drag limit}: In this limit we keep the local velocity $v(u)= \r(u) \o$ constant as we send $\o\to \infty$. Substituting into the radial equation (\ref{rhoeq2}) we see that the equation in this limit becomes  $v'(u)=0$. Therefore the velocity at every point on the string should be the same as its value at the boundary, i.e. $v(u) = L \o \equiv v$. In particular the velocity at the special point $u_c$ should also be the same: $v_c \to v$ in this limit. Inspection at equation
(\ref{el5}) then confirms that energy-loss is indeed dominated by linear drag in this limit:
\be \lab{el6}
\ell^2 \frac{dE}{dt}\to \frac{32 \pi^2 \l_4}{9} \frac{(T\ell)^3}{\Lambda_{QCD}} \frac{v^2}{\sqrt{1-v^2}}.
\ee %
This is the same as the result obtained in \cite{Talavera}, where the drag force acting on a quark in linear motion
with constant velocity $v$ in the $D4/S^1$ plasma was studied. This
conclusion is confirmed by our numerics in figure 11. One observes in
thse plots that the ratio of the total energy loss and the energy loss
due to linear drag becomes 1 for a larger range of $L$ as the
frequency decreases, supporting the argument above.

\begin{figure}[ht]
\lab{LDHigh}
\hspace{-1cm}
\includegraphics[width=2.1in]{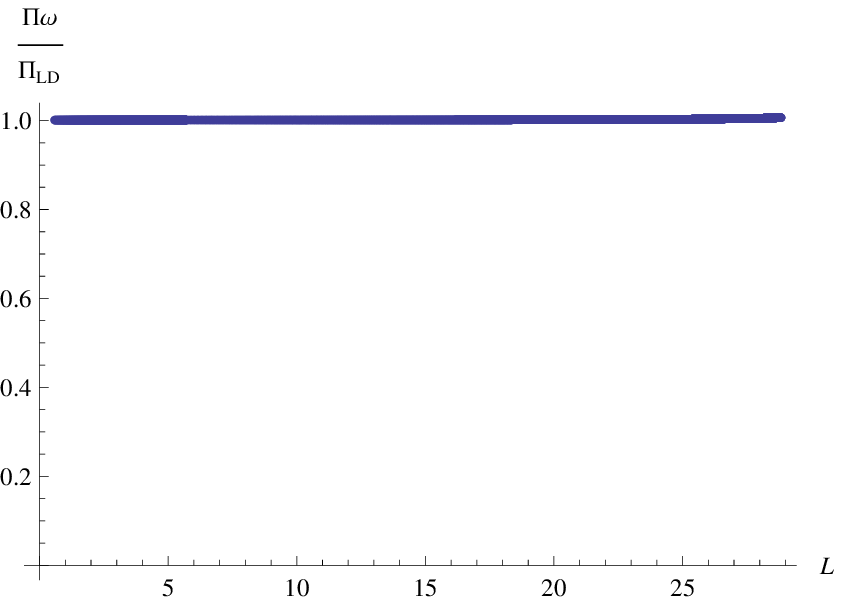}%
\hspace*{.2cm}
\includegraphics[width=2.1in]{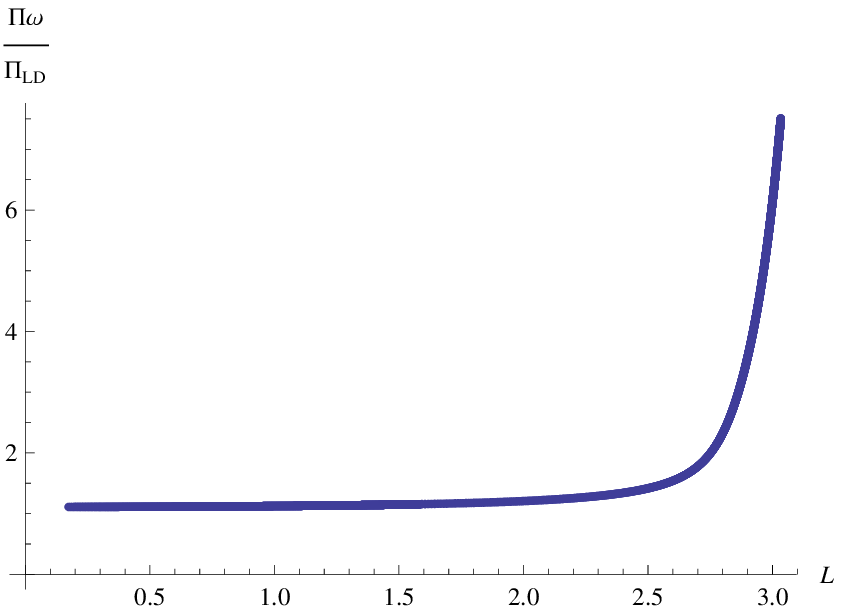}%
\hspace*{.2cm}
\includegraphics[width=2.1in]{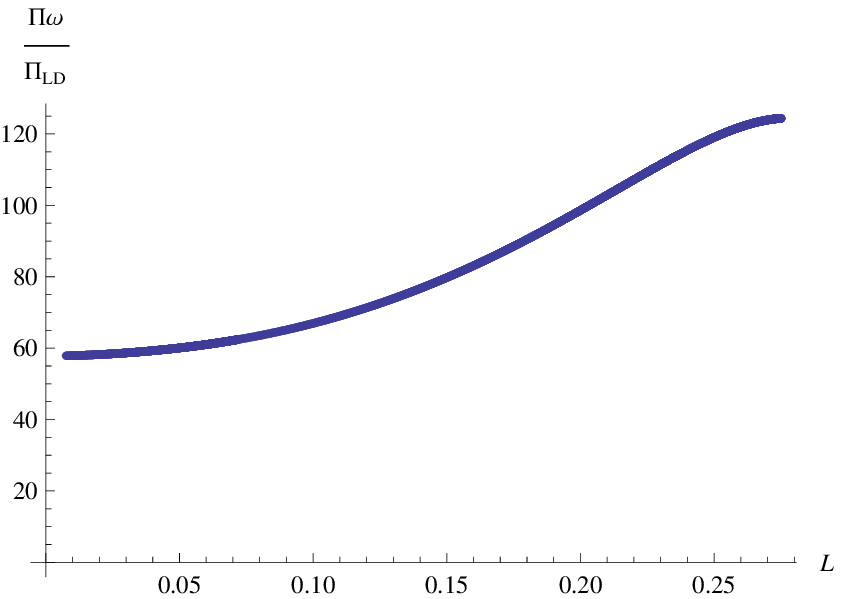}%
\caption{Ratio of the total energy loss and the energy loss due to
  linear drag as a function of the radius of rotation for choices
  $\omega/T $=0.03, 0.3 and 3 (right to left).}
\end{figure}%

\paragraph{Radiation dominated limit:} Now, let us discuss the
opposite limit  $\o\to \infty$, $L\to 0$, $v$ kept constant. In this
limit $\Pi \o$ also diverges\footnote{This can be inferred from the
  fact that in the UV limit the energy loss---that is proportional to
  $\Pi\o$, see eq. (\ref{el2})---should diverge. We also numerically
  checked that, for a fixed v, i.e. $L= v/\o$,  $\Pi \o$ diverges as
  $\o$ increases, see fig. \ref{Khoob}}.
From equations (\ref{uc2}) and (\ref{rc2}) we find that $\r_c\to 1/\o$ in the limit:
\be\lab{uc3}
u_c\to \ell(\Pi \o)^{2/3}, \qquad \r_c\to \frac{1}{\o}\,\, {\textrm as}\,\, \o\to\infty.
\ee
\begin{figure}[ht]
\lab{RadHigh}
\hspace{-1cm}
\includegraphics[width=2.1in]{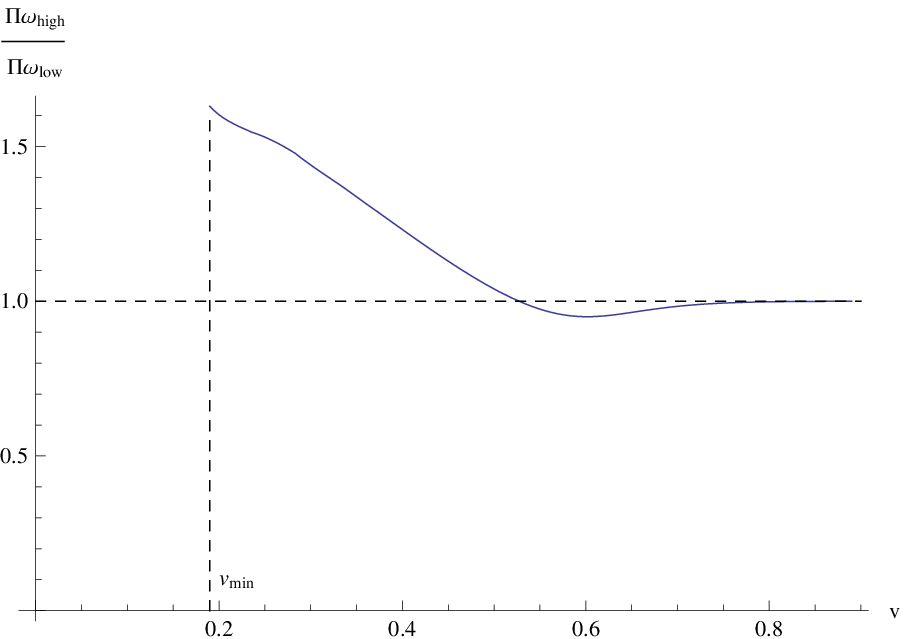} %
\hspace*{.2cm}
\includegraphics[width=2.1in]{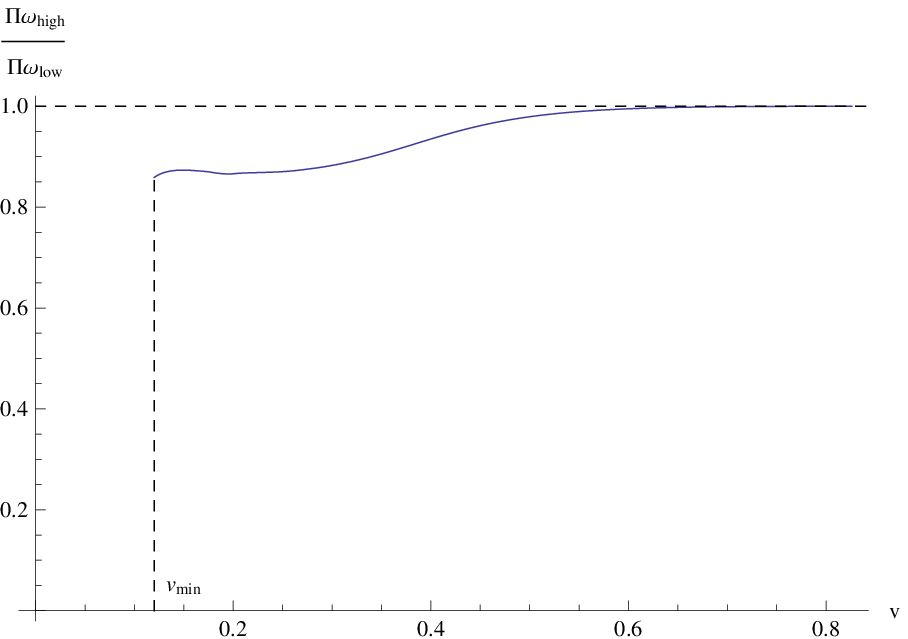}%
\hspace*{.2cm}
\includegraphics[width=2.1in]{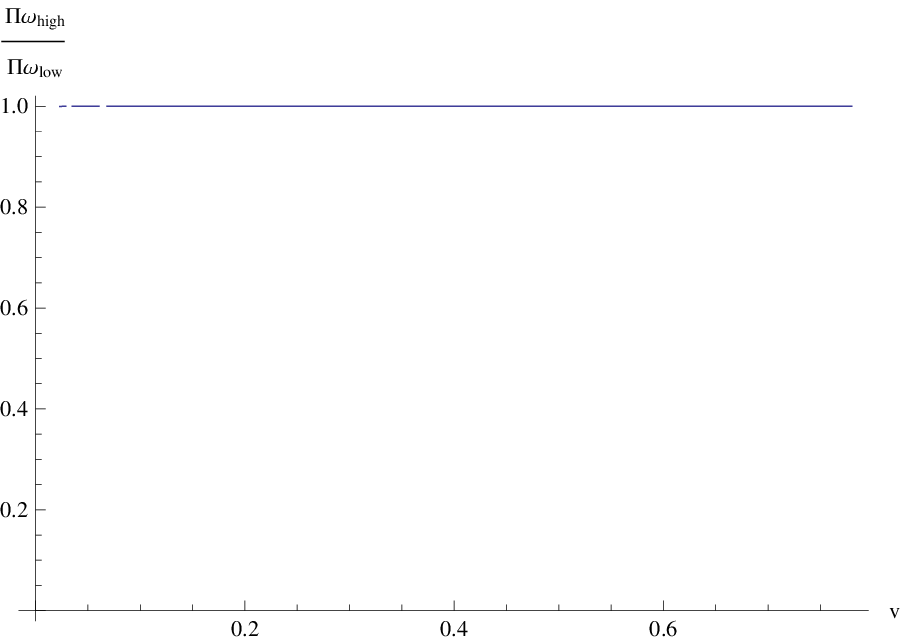}%
\caption{Ratio of the total energy loss and the energy loss due to
  syncrotron radiation as a function of the velocity $v = \o L$ for
  different choices of  $\omega/T $=2.8, 4.6 and 28 (left to
  right). The syncrotron radiation is determined by the low T energy
  loss. We display the minimum allowed value of $v$ in the low T phase by $v_{min}$. }
\end{figure}%

Now, one can easily show that, energy loss of the probe in this limit, is
dominated by  radiation with the following argument: Since $\Pi\o$ becomes large in
this limit, so does $u_c$ according to (\ref{uc3}), therefore $u_c$
comes very close to the boundary. In this part of the solution,
i.e. $u>u_c$, then, the geometry approaches that of
the boundary asymptotics. In particular, this means $f_h\to 1$ in the
limit. The same is true for the low T solution, where again, for very
large $u_c$ the solution in the region $u>u_c$ satisfies $f_k\to 1$.
Therefore both the low T and the high T geometries (\ref{lowT}) and
(\ref{highT}) become the same.
As a result the string Lagrangians and the equations of motion
(\ref{rhoeq1}) and  (\ref{rhoeq2}) also become the same.
Furthermore, comparison of (\ref{uc3}) and (\ref{uc})
shows that the initial condition for solving these equations also
become identical. Therefore, the entire solution and particularly the value of $\rho$ at the boundary should
also become the same in the low and high T phases in the UV
limit. From the general equation (\ref{el2}) then we learn that
\be\lab{UVlim}
\frac{\frac{dE}{dt}\bigg|_{highT}}{ \frac{dE}{dt}\bigg|_{lowT}} \to 1, \qquad {\textrm as}\,\, \o\to\infty.
\ee
The argument we present here supports the other arguments in section
(\ref{lowrad}) that the low T energy loss (\ref{ellow2}) is completely
due to radiation.
%
To confirm the picture we present here, we check  that (\ref{UVlim}) is indeed satisfied by the numerics.
In the figure 12 we compare the low T and high T
energy loss rates at two different frequency $\o=0.3$ and $\o=3$.
We observe that, although the two phases differ substantially for the
smaller frequency,
they become identical (in the range $L > L_{min}$ in the low T phase) as the frequency gets bigger.


\section{Discussion}
\lab{diss}

In this paper, we studied energy loss of probes in uniform circular motion in strongly coupled confining gauge theories. The phenomenon is investigated both in the low and the high T phases of the gauge theory by constructing rotating string solutions in the $D4/S^1$ background. Our findings can be summarized as follows.
\bz
\item There exists a world-sheet horizon in the low T phase for circular motion, as opposed to linear motion. This corresponds to the fact that although there is no drag in the low T phase, there is still energy loss due to syncrotron radiation.
\item Our calculation of the low T energy loss rate determines the Lienard potential for syncrotron radiation
for the $D4/S^1$ theory at strong coupling. We observed that the
simple analytic formula that holds for the ${\cal N}= 4$ theory
\cite{Mikhailov}, that is essentially identical to that of the Lienard
potential for electromagnetic radiation, does not hold here. In particular
it increases with a rate much slower than quadratic in this theory, see figure 10.
This is due to the unrealistic UV completion in the $D4/S^1$ system.
\item We found a lower limit to the energy radiated in the low-T phase, that is proportional to $\Lambda_{QCD}$ corresponding to the fact that at low T, there is no drag but radiation and the latter should be by emission of  glueballs that are gapped. The corresponding statement on the GR side is as follows: There exists an IR cut-off in the geometries dual to confining backgrounds, that is given by $u_k$ in the $D4/S^1$ case, and the regularity of the rotating string solution requires that the world-sheet horizon $u_c> u_k$. This yields a lower bound on the
    world-sheet momentum.
\item Energy loss rate in the high T phase is similar to the conformal
  plasma \cite{Fadafan}. We demonstrated, both analytically and
  numerically, that the energy loss is dominated by syncrotron radiation in the limit $\o\to\infty$, $L\to 0$, $v=\o L$ constant and it becomes identical to that of the low T energy loss formula in this limit.
\item We also showed that the opposite limit $\o\to 0$, $L\to \infty$, $v=\o L$ constant reproduces the linear drag
case studied in \cite{Talavera}.
\ez
All of the results above immediately generalize to arbitrary holographic backgrounds that is dual to a strongly coupled confining gauge theories at large N, where one can neglect $g_s$ and $\alpha'$ corrections.

It will be very interesting to investigate rotating strings in more realistic holographic backgrounds, such as
\cite{GKMN1,Gubser2}. It is also tempting to generalize the study of linearly accelerating probes in \cite{Mexico} to the case of confining gauge theories such as the model we studied here. Finally, it would be very interesting to 
study the energy loss of rotating (as well as in linear motion) hard probes  in the high T phase background in \cite{Takeshi} described by localized D3 branes. 


\section*{Acknowledgments}

We thank F. Ardalan, J. Casalderrey-Solana, E. Kiritsis, M.M. Sheikh-Jabbari,  T. Morita, J. Sonner and U. Wiedemann for interesting and helpful discussions. This research was supported in part by the National Science Foundation under Grant No. NSF PHY05-51164. M. A. would like to thank F. Ardalan and also thanks the CERN Theory Division for its hospitality and financial support.


\end{document}